\documentclass[a4paper]{jpconf}
\usepackage{graphicx}
\usepackage{fancyhdr}
\usepackage{fancybox}

\usepackage{graphicx}
\usepackage{color}
\usepackage{soul}
\usepackage{latexsym}
\usepackage{amssymb}

\def\cta{\cos\theta_A}

\def\mueff{\mu_\mathrm{eff}}

\def\cta{\cos\theta_A}
\def\sta{\sin\theta_A}

\def\tauptaum{\tau^+\tau^-}

\def\sig{\sigma}

\def\ls#1{\ifmath{_{\lower1.5pt\hbox{$\scriptstyle #1$}}}}
\def\lss#1{\ifmath{^{\,\lower2.5pt\hbox{$\scriptstyle #1$}}}}

\def\sig{\sigma}

\def\what{\widehat}

\def\mhi{m_{\hi}}

\def\h{h}

\def\lam{\lambda}
\def\kap{\kappa}
\def\alam{A_\lam}
\def\akap{A_\kap}

\def\wtil{\widetilde}
\def\what{\widehat}
\def\tauptaum{\tau^+\tau^-}

\def\lsim{\mathrel{\raise.3ex\hbox{$<$\kern-.75em\lower1ex\hbox{$\sim$}}}}
\def\gsim{\mathrel{\raise.3ex\hbox{$>$\kern-.75em\lower1ex\hbox{$\sim$}}}}
\def\ifmath#1{\relax\ifmmode #1\else $#1$\fi}

\def\vev#1{\langle #1 \rangle}
\def\lam{\lambda}

\def\mhi{m_{h_1^0}}
\def\etmiss{/ \hskip-8pt E_T}

\def\eg{{\it e.g.}}

\def\msusy{m_{\rm SUSY}}

\def\eg{{\it e.g.}}

\def\hsm{h_{\rm SM}}

\def\hl{h^0}
\def\hh{H^0}
\def\ha{A^0}
\def\hp{h^+}
\def\hm{h^-}

\def\mhh{m_{\hh}}

\def\mhp{m_{\hp}}

\def\tanb{\tan\beta}

\def\mgut{M_U}

\def\cnone{\wt\chi^0_1}

\def\cntwo{\wt\chi^0_2}

\def\mcnone{m_{\cnone}}
\def\mcntwo{m_{\cntwo}}

\def\wt{\widetilde}

\def\cpmone{\wt \chi^{\pm}_1}

\def\mcpmone{m_{\cpmone}}

\def\meff{m_{eff}}

\def\MPL #1 #2 #3 {{\sl Mod.~Phys.~Lett.}~{\bf#1} (#3) #2}
\def\NPB #1 #2 #3 {{\sl Nucl.~Phys.}~{\bf #1} (#3) #2}
\def\PLB #1 #2 #3 {{\sl Phys.~Lett.}~{\bf #1} (#3) #2}
\def\PR #1 #2 #3 {{\sl Phys.~Rep.}~{\bf#1} (#3) #2}
\def\PRD #1 #2 #3 {{\sl Phys.~Rev.}~{\bf #1} (#3) #2}
\def\PRL #1 #2 #3 {{\sl Phys.~Rev.~Lett.}~{\bf#1} (#3) #2}
\def\RMP #1 #2 #3 {{\sl Rev.~Mod.~Phys.}~{\bf#1} (#3) #2}
\def\ZPC #1 #2 #3 {{\sl Z.~Phys.}~{\bf #1} (#3) #2}
\def\IJMP #1 #2 #3 {{\sl Int.~J.~Mod.~Phys.}~{\bf#1} (#3) #2}
\def\NIM #1 #2 #3 {{\sl Nucl.~Inst.~and~Meth.}~{\bf#1} {#3} #2}
\def\lam{\lambda}
\def\br{B}
\def\tauptaum{\tau^+\tau^-}

\def\gam{\gamma}

\def\anti{\overline}
\def\epem{e^+e^-}
\def\mupmum{\mu^+\mu^-}

\def\mupmum{\mu^+\mu^-}

\def\ie{{\it i.e.}}
\def\eg{{\it e.g.}}

\def\anti{\overline}

\def\ai{a_1}
\def\aii{a_2}
\def\mai{m_{\ai}}
\def\maii{m_{\aii}}

\def\fbi{~{\rm fb}^{-1}}

\def\pb{~{\rm pb}}

\def\gev{~{\rm GeV}}
\def\tev{~{\rm TeV}}

\def\hi{\h_1}
\def\hii{\h_2}
\def\hiii{\h_3}
\def\mhi{m_{\hi}}
\def\mhii{m_{\hii}}
\def\mhiii{m_{\hiii}}

\newcommand{\nc}{\newcommand}
\nc{\beq}{\begin{equation}}   \nc{\eeq}{\end{equation}}
\nc{\bea}{\begin{eqnarray}}   \nc{\eea}{\end{eqnarray}}
\nc{\baa}{\begin{array}}      \nc{\eaa}{\end{array}}
\nc{\bit}{\begin{itemize}}    \nc{\eit}{\end{itemize}}
\nc{\ben}{\begin{enumerate}}  \nc{\een}{\end{enumerate}}
\nc{\bce}{\begin{center}}     \nc{\ece}{\end{center}}
\def\beqa{\begin{eqnarray}}
\def\eeqa{\end{eqnarray}}
\def\bed{\begin{description}}
\def\eed{\end{description}}
\def\bmini{\begin{minipage}}
\def\emini{\end{minipage}}
\newcommand{\ba}{\begin{array}}
\newcommand{\ea}{\end{array}}

\def\etmiss{\slash E_T}

\def\eg{{\it e.g.}}

\def\tanb{\tan\beta}

\def\ie{{\it i.e.}}
\def\what{\widehat}

\def\gam{\gamma}

\def\lam{\lambda}
\def\sig{\sigma}

\def\anti{\overline}

\def\vev#1{{\langle #1 \rangle}}

\def\cta{\cos\theta_A}

\def \fbi{~{\rm fb^{-1}}} 
\def\simle{%  ``less than about'' symbol
    \mathrel{\rlap{\raise 0.511ex 
        \hbox{$<$}}{\lower 0.511ex \hbox{$\sim$}}}}

\def\slashchar#1{\setbox0=\hbox{$#1$}           % set a box for #1
   \dimen0=\wd0                                 % and get its size
   \setbox1=\hbox{/} \dimen1=\wd1               % get size of /
   \ifdim\dimen0>\dimen1                        % #1 is bigger
      \rlap{\hbox to \dimen0{\hfil/\hfil}}      % so center / in box
      #1                                        % and print #1
   \else                                        % / is bigger
      \rlap{\hbox to \dimen1{\hfil$#1$\hfil}}   % so center #1
      /                                         % and print /
   \fi}

\def\etmiss{\slashchar{E}_T}

\def\sigsi{\sigma_{SI}}
\def\sigsd{\sigma_{SD}}

\def\cogent{CoGeNT}
\def\asoft{A_{soft}}
\def\caibb{C_{\ai b\anti b}}
\def\caiibb{C_{\aii b\anti b}}
\def\chibb{C_{\hi b\anti b}}
\def\chiibb{C_{\hii b\anti b}}
\def\chiiibb{C_{\hiii b\anti b}}

\begin{document}
\title{The Higgs Sector and \cogent/DAMA-Like Dark Matter in
  Supersymmetric Models}

\author{John F. Gunion}

\address{Department of Physics, UC Davis, Davis CA, 95616}

\ead{jfgunion@ucdavis.edu}

\begin{abstract}
  Recent data from \cogent\ and DAMA are roughly consistent with a
  very light dark matter particle with $m\sim 4-10\gev$ and
  spin-independent cross section of order $\sigsi\sim (1-3)\times
  10^{-4}\pb$.  An important question is whether these observations
  are compatible with supersymmetric models obeying $\Omega h^2\sim
  0.11$ without violating existing collider constraints and precision
  measurements. In this talk, I review the fact the the Minimal
  Supersymmetric Model allows insufficient flexibility to achieve such
  compatibility, basically because of the highly constrained nature of
  the MSSM Higgs sector in relation to LEP limits on Higgs bosons.  I
  then outline the manner in which the more flexible Higgs sectors of
  the Next-to-Minimal Supersymmetric Model and an Extended
  Next-to-Minimal Supersymmetric Model allow large $\sigsi$ and
  $\Omega h^2\sim 0.11$ at low LSP mass without violating LEP,
  Tevatron, BaBar and other experimental limits.  The relationship of
  the required Higgs sectors to the NMSSM ``ideal-Higgs'' scenarios is
  discussed.

\end{abstract}

\section{Introduction}

\cogent~\cite{cogentnew} and DAMA~\cite{DAMAnew} both have hints of
dark matter detection corresponding to a very low mass particle with
very large spin-independent cross section. In~\cite{Hooper:2010uy}, it
is claimed that a consistent explanation for both hints is provided if
$\sigsi\sim (1.4-3.5)\times 10^{-4}\pb$, for $m_{DM}=(9-6)\gev$. Of
course, the required $\sigsi$ is dependent on the assumed local relic
density and for example is reduced by $\sim 60\%$ if
$\rho=0.485\gev/{\rm cm}^3$~\cite{Pato:2010yq} is employed rather than
the usual $\rho=0.3\gev/{\rm cm}^3$.

One would hope that \cogent/DAMA-like $\sigsi$ at low $m_{DM}$ could
be consistent with simple supersymmetric models. However, the MSSM
fails.  For low $\mcnone$, the MSSM generically predicts a value for
$\Omega_{\cnone} h^2$ that is far above the observed value; it is only
for extreme choices that one can achieve $\Omega_{\cnone} h^2\sim
0.11$~\cite{wmap}. For these same choices it turns out that $\sigsi$
takes on its maximum possible value of $\sim 0.17\times
10^{-4}\pb$. This maximum value of $\sigsi$ can be understood as
follows. The cross
section for $\cnone$-nucleon scattering is dominated by CP-even Higgs
exchange and is given approximately by
\begin{eqnarray}
\sigsi
&\approx& 0.17 \times 10^{-4}\pb \, \bigg(\frac{N^2_{13}}{0.1}\bigg) \bigg(\frac{\tan \beta}{50}\bigg)^2 \bigg(\frac{100 {\rm GeV}}{\mhh}\bigg)^4\cos^4\alpha\,,
\end{eqnarray}
where we have written $\cnone=N_{11}\wtil B+ N_{12} \wtil
W^3+N_{13}\wtil H_d+N_{14}\wtil H_u$ and referenced the most
optimistic values of $N_{13}^2$, $\tanb$ and $\mhh$.  In the above,
$N_{13}^2$ cannot be much larger than $0.1$ because of limits on the
$Z$ invisible width, $\tanb>50$ enters a non-perturbative Yukawa
coupling domain and $\mhh<100\gev$ is not allowed by LEP limits.
Further, to achieve even this most maximal value of $\sigsi$ in the
MSSM, one must ignore the Tevatron limit, $\br(B_s\to \mupmum)\leq
5.8\times 10^{-8}$.  Once imposed, the largest $\sigsi$ for scenarios
with $\mcnone$ in the \cogent/DAMA region and with $\Omega_{\cnone}
h^2\sim 0.1$ is $\sigsi\sim 0.017\times
10^{-4}\pb$~\cite{Feldman:2010ke}, a factor of roughly 100 below the
$\sigsi$ needed to explain the \cogent/DAMA events.

Thus, it is natural to turn to the even more attractive NMSSM model.
The NMSSM is defined by adding a single SM-singlet superfield $\what
S$ to the MSSM and imposing a $Z_3$ symmetry on the superpotential,
implying 
\beq \label{1.1} 
W=\lambda \ \widehat{S} \widehat{H}_u
\widehat{H}_d + \frac{\kappa}{3} \ \widehat{S}^3 
\eeq
 The reason for
imposing the $Z_3$ symmetry is that then only dimensionless couplings
$\lambda$, $\kappa$ enter. All dimensionful parameters will then be
determined by the soft-SUSY-breaking parameters.  In particular, the
$\mu$ problem is solved via $ \mueff=\lam \vev S$ for which $\mueff$
is automatically of order a $\tev$ (as required) since $\vev{S}$ is of
order the SUSY-breaking scale, $\msusy$, which will be below a $\tev$.

The extra singlet field $\what S$ implies: $5$ neutralinos,
  $\wtil\chi^0_{1-5}$ with $\cnone=N_{11}\wtil B+ N_{12} \wtil
W^3+N_{13}\wtil H_d+N_{14}\wtil H_u+N_{15}\wtil S$ being either singlet or bino,
  depending on $M_1$; $3$
  CP-even Higgs bosons, $\hi,\hii,\hiii$; and $2$ CP-odd Higgs bosons,
  $\ai,\aii$. 

  The soft-SUSY-breaking terms corresponding to the terms in $W$ are:
  \beq 
\lambda A_{\lambda} S H_u H_d + \frac{\kappa} {3} A_\kappa
  S^3\,.  
\eeq 
It is important to recall that when $\alam,\akap\to 0$,
  the NMSSM has an additional $U(1)_R$ symmetry, in which limit the
  $\ai$ is pure singlet and $\mai=0$.
If, $\alam,\akap=0$ at $\mgut$, RGE's give $\alam\sim 100\gev$ and
$\akap\sim 1-20\gev$, resulting in $\mai<2m_B$ (see later) being quite
natural and not fine-tuned~\cite{Dermisek:2006wr}. In this situation
$\ai$ is still primarily singlet so that $\cta$ as defined by
$\ai=\cta A_{MSSM}+\sta A_S$  is typically quite
small.~\footnote{Here, $A_{MSSM}$ is the usual doublet
pseudoscalar of the MSSM two-doublet Higgs sector and $A_S$ is the
CP-odd component of the complex scalar field residing in the singlet
superfield.} A light singlet-like $\ai$ also arises in the $U(1)_{PQ}$
symmetry limit of $\kap,\kap\akap=0$.

As is well known, the NMSSM maintains all the attractive features
(especially coupling constant unification and automatic
electroweak symmetry breaking from renormalization group evolution of
the soft SUSY-breaking stop masses) of the MSSM while avoiding
important MSSM problems. In particular, the level of finetuning is
greatly reduced in ``ideal Higgs'' scenarios~\cite{Dermisek:2005ar} in
which the $\hi$ has SM-like $WW,ZZ$ couplings and $\mhi\lsim 105\gev$
but escapes LEP limits via $\hi\to\ai\ai\to 4\tau$
($\mai<2m_B$). Further, $\mhi\lsim 105\gev$ implies excellent precision
electroweak consistency and suitably strong baryogenesis. In addition,
the long-standing LEP excess in the $Z+b\anti b$ final state
near $M_{b\anti b}\sim 100\gev$ is well fit if $\mhi$ is in the
vicinity of $100\gev$ and $\br(\hi\to b\anti b)\sim 0.1-0.25$, the
latter being automatic when $\br(\hi\to\ai\ai)\sim 0.75-0.9$.

However, if $\hi$ is SM-like then the arguments regarding limitations
on achieving large $\sigsi$ given above continue to apply.  An
alternative yielding much larger maximum $\sigsi$~\cite{Gunion:2010dy}
is to arrange for the lightest Higgs, $\hi$, to have enhanced coupling
to down-type quarks while it is the $\hii$ that couples to $WW,ZZ$ in
SM-like fashion. We term this kind of scenario an ``inverted Higgs''
(IH) scenario.  Many large $\sigsi$ scenarios have $\mhi<90\gev$ and
$\mhii\lsim 110\gev$, and are thus still pretty ideal in the sense
described in the previous paragraph. We call such scenarios ``inverted
ideal Higgs'' (IIH) scenarios. In the general NMSSM context, it is
straightforward~\cite{gunion} to adjust $\mai$ so as to obtain
$\Omega_{\cnone} h^2\sim 0.1$ (using $\cnone\cnone\to \ai \to X$ with
$\mai$ small). Further, in IH and IIH scenarios we
found~\cite{Gunion:2010dy} that one can achieve $\sigsi\sim
(0.1-0.2)\times 10^{-4}\pb$ {\it without violating the $\br(B_s\to
  \mupmum)$ bound, or any other bound.}  But, to get $\sigsi$ as large
as $1\times 10^{-4}$ requires violating $(g-2)_\mu$ quite badly, and
having some enhancement of the $s$-quark content of the nucleon.

In a second paper~\cite{Belikov:2010yi}, we showed that an extended
version of the NMSSM would allow $\sigsi$ as large as needed for the
\cogent/DAMA events while maintaining consistency with all
constraints, including $\Omega_{\cnone} h^2\sim 0.11$. In particular, we
considered the ENMSSM in which we only generalize the superpotential
and soft-SUSY-breaking potential, keeping to just one singlet
superfield. The extended superpotential is given by
\begin{equation}
v_0^2 \hat{S} + \frac{1}{2} \mu_S \hat{S}^2 + \mu \hat{H}_u \hat{H}_d + \lambda \hat{S} \hat{H}_u \hat{H}_d
+ \frac{1}{3} \kappa \hat{S}^3 ~,
\label{eq:W}
\end{equation}
and the soft Lagrangian is
\begin{eqnarray}
B_{\mu} H_u H_d+\frac{1}{2} m_S^2 |S|^2 + B_S S^2+ \lambda A_\lambda S
H_u H_d +  \frac 1 3\kappa A_\kappa S^3 + H.c.
\label{eq:Lsoft}
\end{eqnarray}
Note that the presence of explicit $\mu$ and $B_\mu$ terms. 
These reduce the appeal of the model somewhat, but there are
string-theory-inspired sources for such explicit terms.  We found that
scenarios in the ENMSSM with the largest $\sigsi$ are ones in which
the $\cnone$ is singlino-like and the $\hi$ is  largely singlet
(rather than bino-like and mainly $H_d$, respectively, as in the IH NMSSM
scenarios).  To first approximation, $\Omega_{\cnone} h^2$ is controlled by
$\cnone\cnone\to \hi\to X$ and $\sigsi$ is determined by $\hi$
exchange between the $\cnone$ and the down-type quarks in the nucleon,
especially $s$ and $b$. We term this scenario the singlino-singlet (SS)
scenario.

In a very recent paper~\cite{Draper:2010ew}, it is found that the SS
type scenario can be realized in the NMSSM provided the $\hi$ has
$\mhi\lsim 1\gev$. They call their scenario the Dark Light Higgs (DLH)
scenario. This scenario requires a considerable degree of finetuning
for the couplings, but is consistent with current experimental constraints.

While we await confirmation of the \cogent/DAMA excesses, it is very
interesting to consider the implications of all the above scenarios
for Higgs physics at the LHC.  That will be the topic of the remainder
of this presentation.

\section{The Inverted Higgs Scenarios}

The largest elastic scattering cross sections arise in the case of
large $\tan \beta$, significant $N_{13}$ (the Higgsino component of
the $\cnone$), and relatively light $m_{H_d}$,
where $H_d$ is the Higgs with enhanced coupling to down quarks,
$C_{H_d d\anti d}\sim \tanb$.
 In this limit, the relevant scattering amplitude is
\begin{equation}
\frac{a_d}{m_d}\approx \frac{-g_2 g_1 N_{13}N_{11} \tan \beta}{4 m_W m^2_{H_d}},
\end{equation}
which in turn yields
%%
%\begin{equation}
%\frac{f_{p,n}}{m_{p,n}} \approx [f^{(p,n)}_{T_s}+\frac{2}{27}f^{(p,n)}_{TG}]  \times [\frac{-g_2 g_1 N_{13}N_{11} \tan \beta}{4 m_W m^2_{H_d}}]
%\end{equation}
%%
\begin{eqnarray}
\sigsi &\approx& \frac{g^2_2 \, g^2_1 \, N^2_{13}N^2_{11} \, \tan^2 \beta \, m^2_{\cnone} \, m^4_{p,n}}{4\pi \, m^2_W m^4_{H_d} \, (m_{\cnone}+m_{p,n})^2} \bigg[f^{(p,n)}_{T_s}+\frac{2}{27}f^{(p,n)}_{TG}\bigg]^2 \nonumber \\
&\approx& 1.7 \times 10^{-5}\pb \, \bigg(\frac{N^2_{13}}{0.10}\bigg) \bigg(\frac{\tan \beta}{50}\bigg)^2 \bigg(\frac{100 {\rm GeV}}{m_{H_d}}\bigg)^4.
\end{eqnarray}
Constraints on the light $\hi\sim H_d$ configuration are
significant. We had to update
NMHDECAY~\cite{Ellwanger:2004xm,Ellwanger:2005dv} to include all the
latest constraints. We then linked to
micrOMEGAs~\cite{Belanger:2006is} for the computation of
$\Omega_{\cnone}h^2$ and $\sigsi$ as in
NMSSMTools~\cite{nmssmtools}. The important constraints are:
\ben
\item Constraints on the neutral Higgs sector from $Z\hii$ at LEP.
  These are important since $\msusy$ should be low in order to
  minimize $\mhi$ and this keeps $\mhii$ low.  In these cases, the
  $\hii$ can be in the ``ideal'' zone ($\mhii<105\gev$) and escapes
  LEP detection via $\hii\to \ai\ai\to 4\tau$ decays with $\mai<2m_B$
  (but close to $2m_B$ in order to avoid BaBar limits on
  $\Upsilon_{3S}\to \gam \ai\to \gam \tauptaum$). Of course, we also
  require compliance with the recent ALEPH limits~\cite{Schael:2010aw}
  on the $\epem\to Z 4\tau$ channel.

\item LEP constraints on $\hi\ai$ and $\hi\aii$.
  The $\hi\ai$ cross section is $\propto maximal\times
  (\cta)^2$. Thus, small $\cta$ is desirable, which fits with both the 
  approximate $U(1)_R$ limit mentioned earlier and the need
  to not have overly strong $\cnone\cnone\to \ai^*\to X$
  annihilations (as required to achieve adequate $\Omega_{\cnone}
  h^2$) when $\mai$ is small and, in particular, not far from the
  $2\mcnone\sim \mai$ resonance pole. 

\item Tevatron direct Higgs production limits.  There are two
  especially relevant limits given the need to focus on large $\tanb$
  in order to achieve large $\sigsi$. The first is the limit on
  $b\anti b \hi$ with $\hi\to\tauptaum$ associated
  production~\cite{Benjamin:2010xb}, which scales as $C_{\hi b\anti
    b}^2\sim \tan^2\beta$, the latter being something we want to
  maximize. The second limits are those on $t\to h^+ b$ with $h^+\to
  \tau^+\nu_\tau$~\cite{:2009zh} (dominant at large $\tanb$). These
  are critical to include since the $h^+$ tends to be quite light
  (e.g. $\sim 120-140\gev$) when $\hi$ is $H_d$-like and the $\hii$ is
  SM-like.

\item Limits from $\Upsilon$ decays.  These are especially crucial for
  constraining scenarios with $\mai<2m_B$. We have included the latest
  $\Upsilon_{3S}\to \gam \mupmum,\gam\tauptaum$
  limits~\cite{Aubert:2009cp,Aubert:2009cka}, whose impacts on ideal
  Higgs scenarios were explored in \cite{Dermisek:2010mg}.

\item $B$-physics constraints.  The most restricting constraint arises
  from the very strong Tevatron limit of $\br(B_s\to
  \mupmum)<5.8\times 10^{-8}$. At large
  $\tanb$, achieving a small enough value fixes $A_t$ as a function of
  $\msusy$. Next comes $b\to s\gam$.  The $\mueff>0$ scenarios have
  roughly $1\sigma$ discrepancy with the $2\sigma$ experimental
  window.  In contrast, the $\mueff<0$ scenarios only rarely have a
  $b\to s\gam$ problem.  One must also check $B^+\to
  \tau^+\nu_\tau$. $\mueff>0$ scenarios pass easily, but $\mueff<0$
  scenarios with the largest $\sigsi$ have $1\sigma-2\sigma$ deviations from
  the experimental $2\sigma$ window.

\item $(g-2)_\mu$. This is possibly crucial.  For $\mueff<0$, the
  largest $\sig_{SI}$ values are achieved when $(g-2)_\mu$ is a few
  sigma outside the $2\sigma$ limits including theoretical
  uncertainties.  If $(g-2)_\mu$ is strictly required to lie within
  the $2\sigma$ window, then the largest $\sigsi$ that can be achieved
  for $\mueff<0$ is about a factor of 50 below that needed for the
  \cogent/DAMA events.  For $\mueff>0$, the largest $\sig_{SI}$ points
  yield $(g-2)_\mu$ within the $2\sigma$ exp.+theor. window, but after
  including all other constraints the $\sig_{SI}$ values for
  $\mueff>0$ are not as large as those found with $\mueff<0$ (before
  imposing the $(g-2)_\mu$ constraint).

\item $\Omega_{\cnone} h^2$: Of course, we require that any accepted scenario
  have correct relic density within the 
  experimental limits encoded in NMSSMTools, $0.094\leq \Omega_{\cnone} h^2\leq 0.136$.

\een

\begin{figure}[h!]
\begin{center}
\vspace*{-.5in}
\hspace*{-.3in}
\includegraphics[width=0.5\textwidth,angle=90]{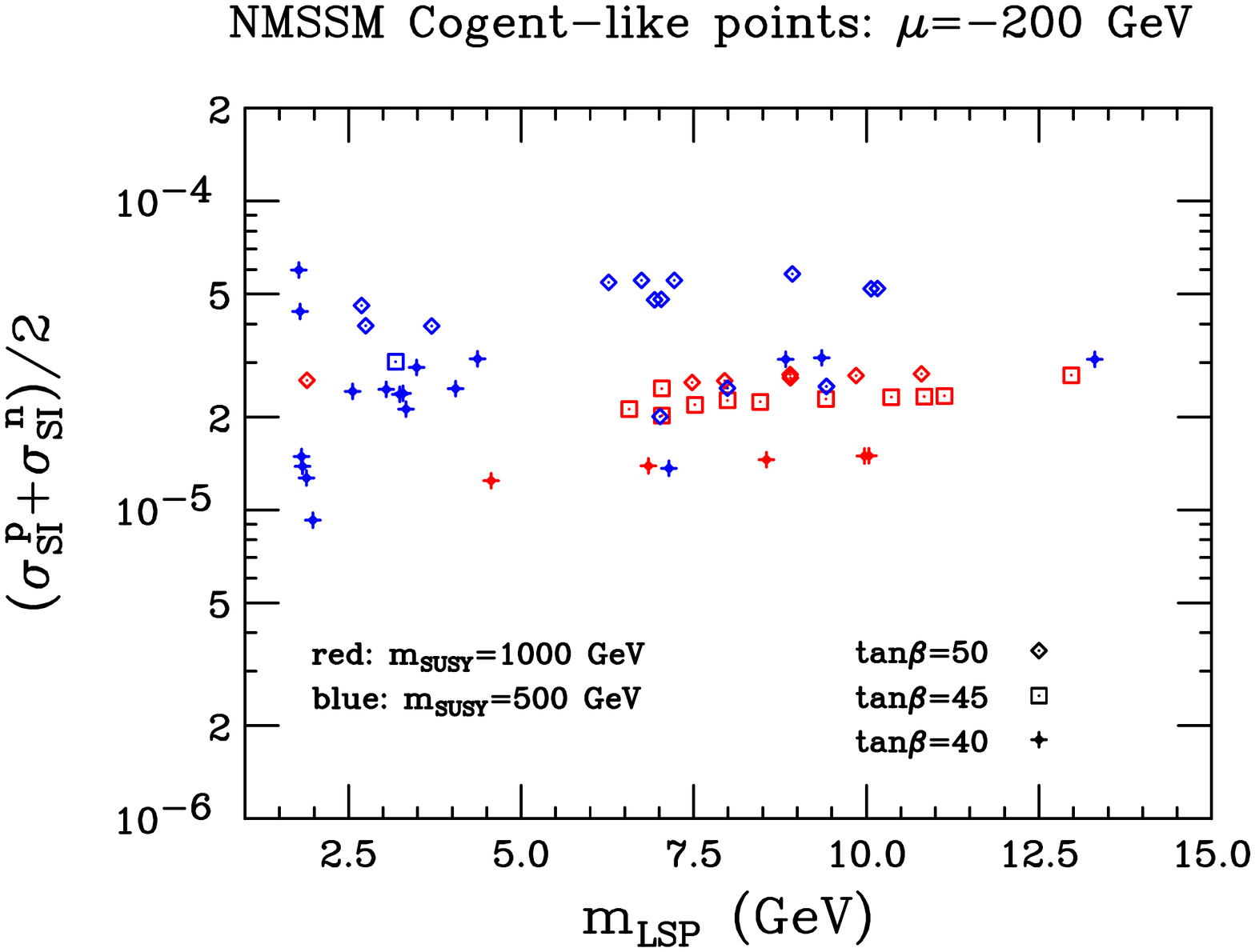}\hspace*{-.8in}\includegraphics[width=0.5\textwidth,angle=90]{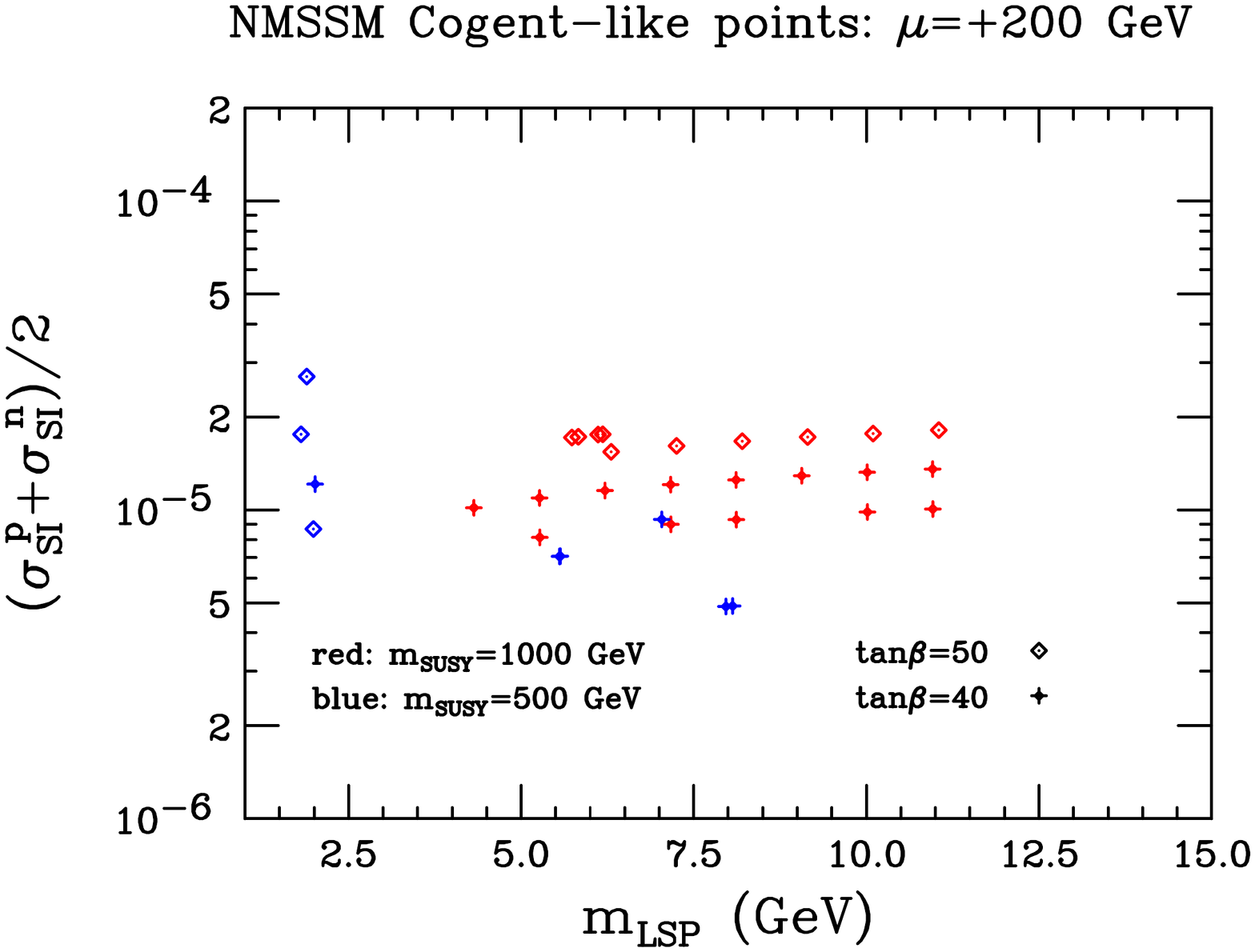}
\end{center}
\vspace*{-.5in}
\caption{ All points obtained after imposing level-I constraints (see
  text) but without imposing either Tevatron direct Higgs
  production limits or 
  $(g-2)_\mu$ limits. }
\label{allpts}
\end{figure}

\begin{figure}[h!]
\begin{center}
\vspace*{-.6in}
\hspace*{-.3in}\includegraphics[width=0.5\textwidth,angle=90]{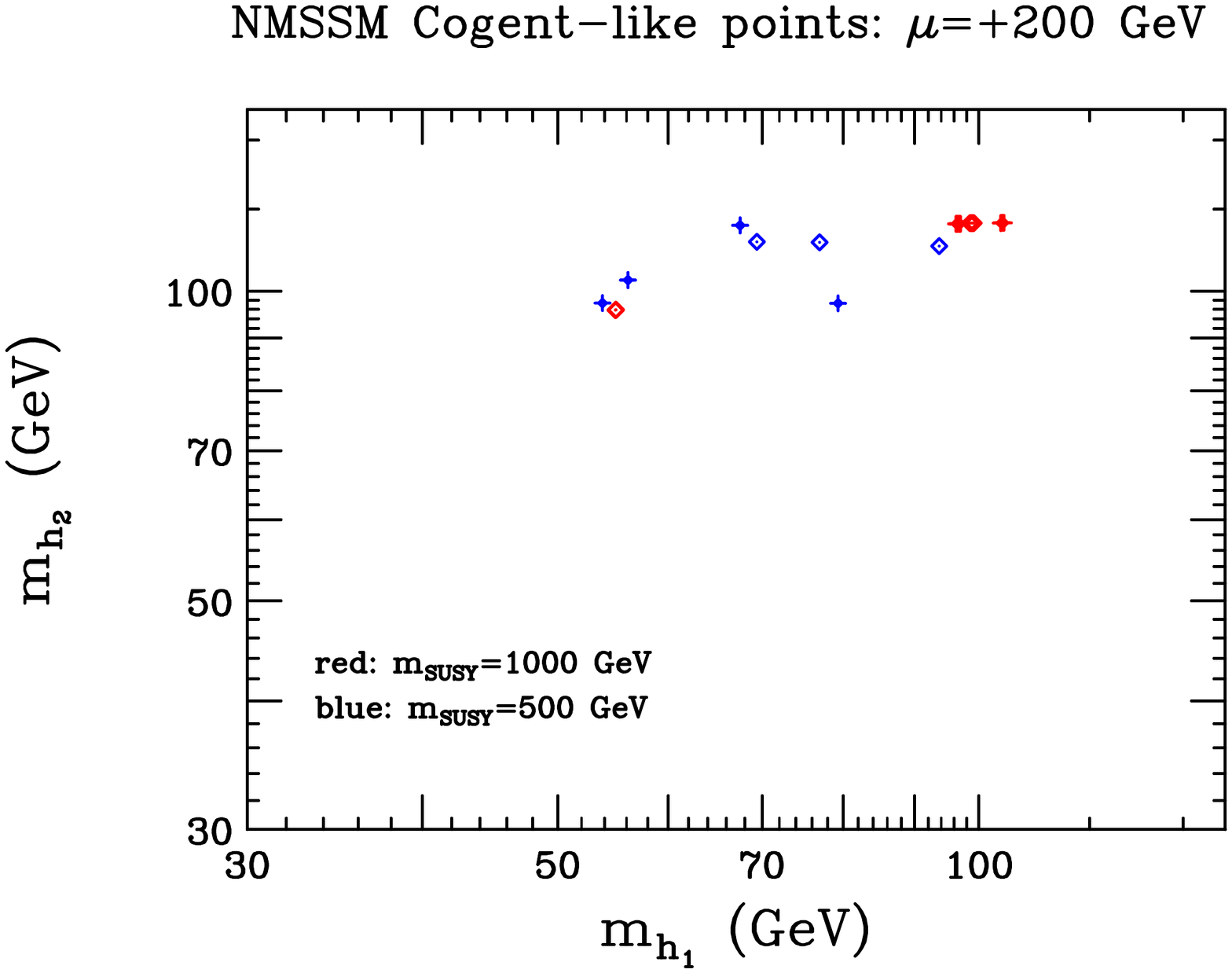}\hspace*{-.8in}\includegraphics[width=0.5\textwidth,angle=90]{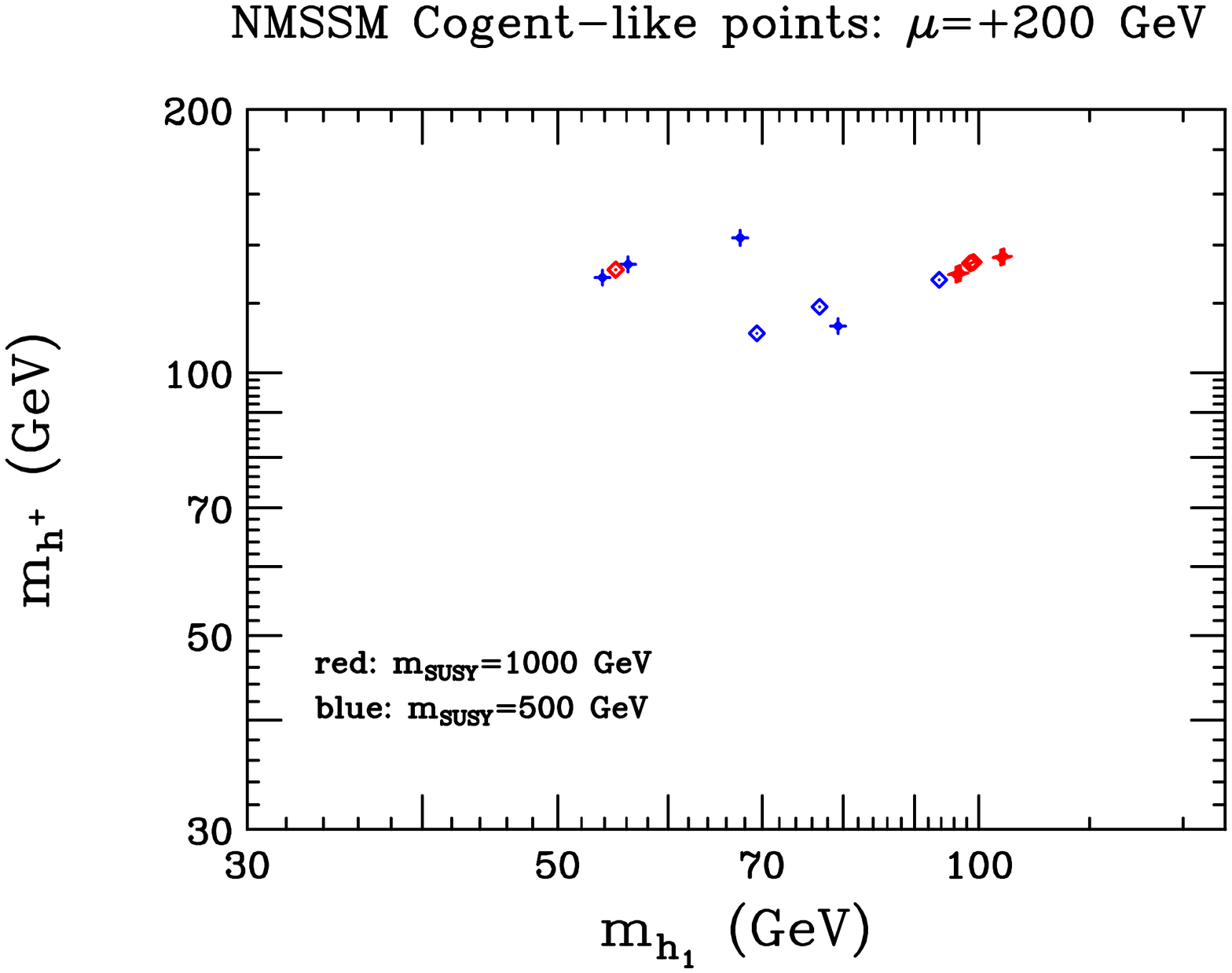}
\end{center}
\vspace*{-.5in}
\caption{ $\mhii$ and $\mhp$ vs. $\mhi$ for $\mueff=+200\gev$ points.
  Only level-I (see text) constraints are
  imposed. There is a great amount of point overlap in these plots.}
\label{massesmu+200}
\end{figure}

\begin{figure}[h!]
\begin{center}
\vspace*{-.6in}
\hspace*{-.3in}\includegraphics[width=0.5\textwidth,angle=90]{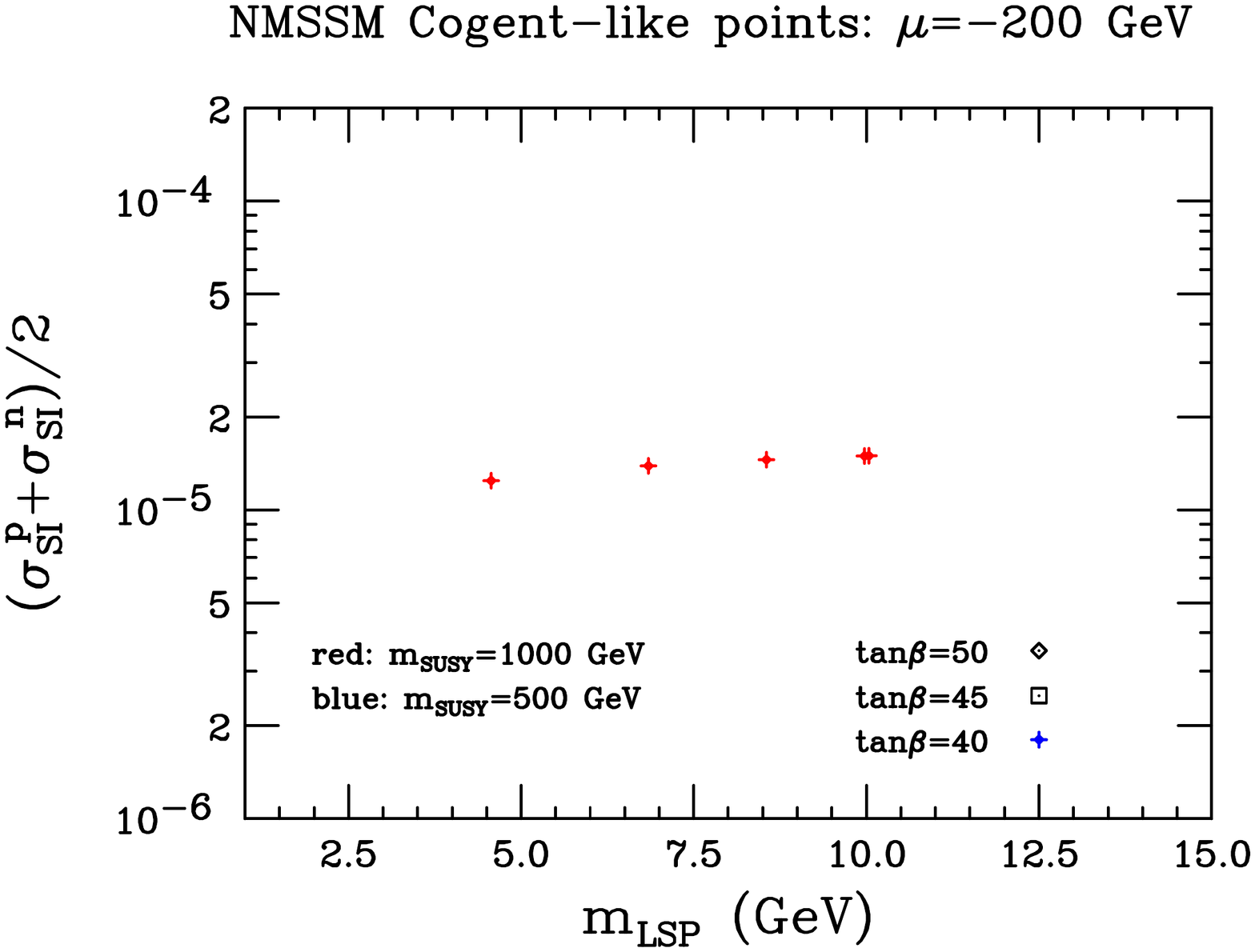}\hspace*{-.8in}\includegraphics[width=0.5\textwidth,angle=90]{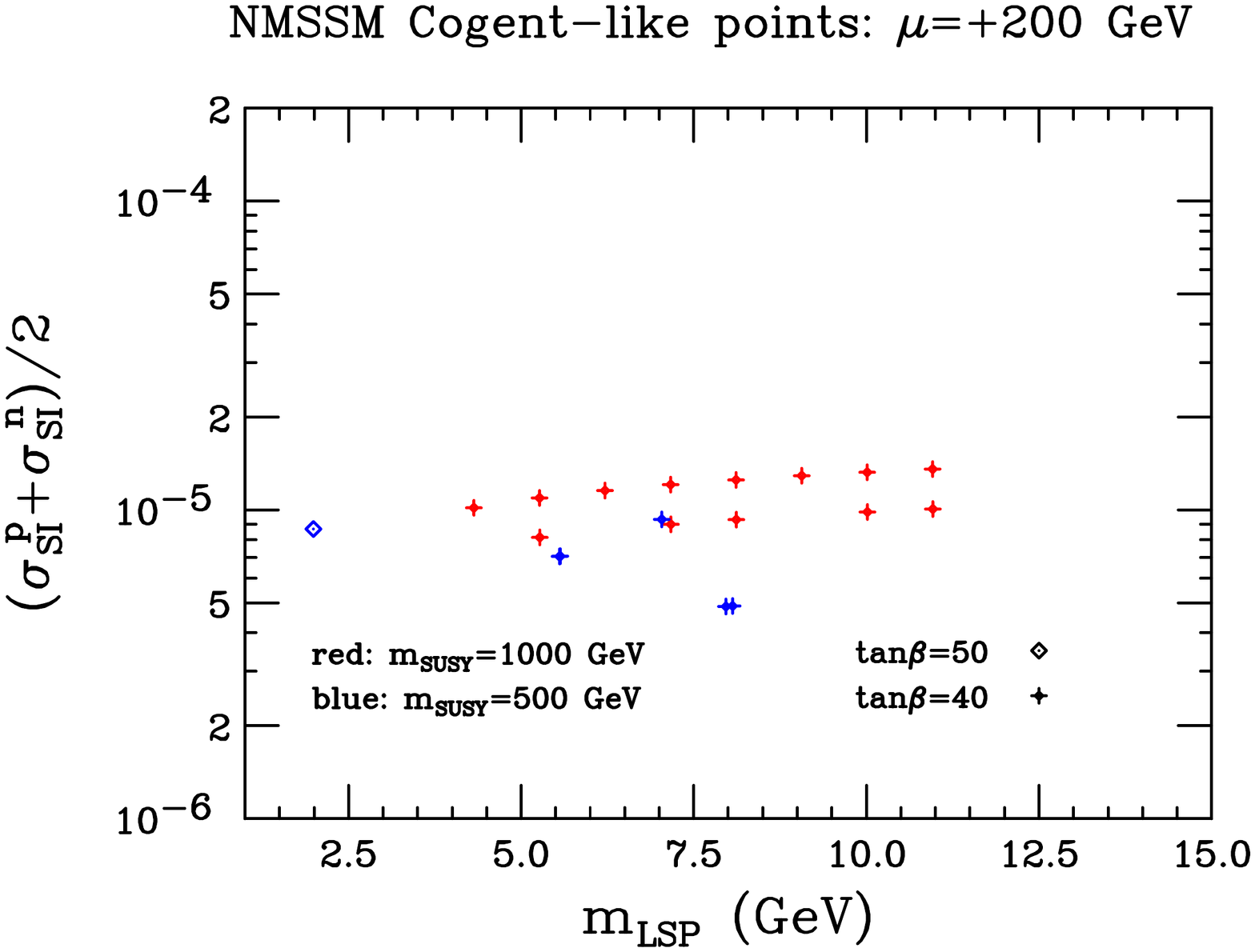}
\end{center}
\vspace*{-.5in}
\caption{ $\sigsi$ vs. $\mcnone$ for points fully consistent with Tevatron
  limits on $b\anti b +Higgs$ and $t\to \hp b$. Level-I constraints
  are imposed. $(g-2)_\mu$ is still terrible (perfectly ok) for the
  surviving $\mueff<0$ ($\mueff>0$) points.}
\label{tevok}
\end{figure}

To illustrate the impact of some of the constraints discussed above, I
give three figures.  In the first, Fig.~\ref{allpts}, representative
maximal values of $\sigsi$ (averaged over protons and neutrons) are
plotted after: a) imposing LEP limits; b) requiring
$0.094<\Omega_{\cnone} h^2<0.136$; c) imposing BaBar limits from
$\Upsilon_{3S}$ decays; and d) requiring $\br(B_s\to
\mupmum)<5.8\times 10^{-8}$.  These are termed ``level I''
constraints.  Note that $\sigsi\sim 6\times 10^{-5}\pb$ ($2\times
10^{-5}\pb$) can be achieved for $\mueff<0$ ($\mueff>0$). In
Fig.~\ref{massesmu+200}, the masses of the CP-even Higgs bosons are
plotted for the $\mueff>0$ cases.  Note how low all these masses are.
These same plots for $\mueff<0$ would be quite similar.

The low masses imply that Tevatron limits on direct Higgs production
could be important.  In Fig.~\ref{tevok}, we plot the points from
Fig.~\ref{allpts} that are fully consistent with Tevatron limits on
$b\anti b + Higgs$ with $Higgs\to \tauptaum$ and on $t\to \hp b$ with
$\hp\to \tau^+\nu_\tau$. Typically, one finds  maximal $\sigsi$ values in the
$(1-2)\times 10^{-5}\pb$ range (a factor at least $5$ below the $\sigsi$
needed for \cogent/DAMA). Many of the points of Fig.~\ref{allpts} with
larger $\sigsi$ would survive if we relaxed these direct production
limits by $1\sigma$ (combined experimental and theoretical error). 
The points of the $\mueff<0$ plot in Fig.~\ref{tevok} would all be
eliminated if the predicted $(g-2)_\mu$ is required to be within
$2\sigma$ of the observed value, whereas all the $\mueff>0$ points are
very consistent with the observed $(g-2)_\mu$.

{\tiny 
\begin{table}[t!]
  \caption{\small Properties of a particularly attractive but
    phenomenologically complex NMSSM point with
    $\mueff=+200\gev$, $\tanb=40$ and $\msusy=500\gev$. All Tevatron
    limits ok. $\hiii$ is the most SM-like. In the last row, the brackets
    give the range of $B$-physics predictions for this point after including
    theoretical errors as employed in NMHDECAY. \label{pt10p}}
\vspace*{.05in}
\begin{center}
\begin{tabular}{|c|c|c|c|c|c|c|c|}
\hline
$\lam$ & $\kap$ & $\alam$ & $\akap$ & $M_1$ & $M_2$ & $M_3$ & $\asoft$
\cr
\hline
$0.081$ &   $0.01605$ & $  -36 \gev$ & $-3.25\gev $ & $8\gev$ &
$200\gev$ & $300\gev$ & $479\gev$ \cr
\hline
\end{tabular}

\begin{tabular}{|c|c|c|c|c|c|}
\hline
$\mhi$ & $\mhii$ & $\mhiii$ & $\mai$ & $\maii$ & $\mhp$ \cr
\hline
$53.8\gev$ & $97.3\gev$ & $126.2\gev$ & $10.5\gev$ & $98.9\gev$ &
$128.4\gev$  \cr 
\hline
\end{tabular}

\begin{tabular}{|c|c|c|c|}
\hline
 $C_V(\hi)$ & $C_V(\hii)$ & $C_V(\hiii)$ & $\meff$ \cr
\hline
$-0.505$ & $0.137$ & $0.852$ & $101\gev$ \cr
\hline
\end{tabular}
\begin{tabular}{|c|c|c|c|c|}
\hline
$\chibb$ & $\chiibb$ & $\chiiibb$ & $\caibb$ & $\caiibb$ \cr
\hline
$0.24$ & $39.7$ & $-5.1$ & $6.7$ & $39.4$ \cr
\hline
\end{tabular}

\begin{tabular}{|c|c|c|c|c|c|c|c|}
\hline
$\mcnone$ & $N_{11}$ & $N_{13}$ & $\mcntwo$ & $\mcpmone$ & $\sigsi$ & $\sigsd$ & $\Omega_{\cnone} h^2$  \cr 
\hline
$7\gev$ & $-0.976$ & $-0.212$ & $79.1\gev$  & $153\gev$ & $0.93 \times 10^{-5} \pb$ & $0.45\times 10^{-4}\pb$ & $0.12$ \cr
\hline
\end{tabular}

\begin{tabular}{|c|c|c|c|}
\hline  
$\br(\hi\to\ai\ai)$ & $\br(\hii\to 2b,2\tau)$ & $\br(\hiii\to 2h+2a)$
& $\br(\hiii\to 2b,2\tau)$ \cr 
\hline
   $0.96$ & $0.87,0.12$ & $0.3$  & $0.58,0.09$  \cr
  \hline
\end{tabular}

\begin{tabular}{|c|c|c|c|c|}
  \hline
$\br(\ai\to  jj)$ & $\br(\ai\to 2\tau)$ & $\br(\ai\to 2\mu)$ &
$\br(\aii\to 2b,2\tau)$ & $\br(\hp\to \tau^+\nu)$ \cr
\hline
$0.28$ & $0.79$ & $0.003$ & $0.87,0.12$ & $0.97$ \cr
\hline
\end{tabular}

\begin{tabular}{|c|c|c|c|}
\hline
$\br(B_s\to\mupmum)$ & $\br(b\to s\gam)$ & $\br(\hp\to \tau^+\nu_\tau)$
& $(g-2)_\mu$ \cr
\hline
$[1.7-6.0]\times 10^{-9}$ & $[5.8-12.5]\times  10^{-4}$ & $[0.91-4.22]\times 10^{-4}$ &
$[4.42-5.53]\times 10^{-9}$ \cr 
\hline
\end{tabular}
\end{center}
\caption{\small The $\pm 2\sigma$ experimental ranges for the $B$ physics
  observables tabulated in the last row of
  Table~\ref{pt10p}. \label{explimits}}
\begin{center}
\begin{tabular}{|c|c|c|c|}
\hline
$\br(B_s\to\mupmum)$ & $\br(b\to s\gam)$ & $\br(\hp\to \tau^+\nu_\tau)$
& $(g-2)_\mu$ \cr
\hline
$<5.8\times 10^{-8}$ (95\% CL) & $[3.03-4.01]\times  10^{-4}$ & $[0.34-2.3]\times 10^{-4}$ &
$[0.88-4.6]\times 10^{-9}$ \cr 
\hline
\end{tabular}
\end{center}
\caption{\small LHC Neutral Higgs Discovery Channels ($b\anti
  b\hii,b\anti b\aii\to
b\anti b 2\tau$ absent since $\mhii\sim\maii<100\gev$, the lower limit of
the studies used --- this should be a highly viable mode) (also
$t\anti t \to b \anti t \hp\to\tau^+\nu X$ = excellent channel at LHC)}
\begin{center}
\begin{tabular}{|c|c|c|c|c|}
\hline
$L=30\fbi$ & \multicolumn{4}{c|}{$L=300\fbi$} \\
\hline
$WW\to \hiii\to 2\tau$ & $b\anti b \hiii\to b\anti b 2\tau$ &
$gg\to\hiii\to 4\ell$ & $gg\to\hiii\to 2\ell 2\nu$ & $WW\to\hiii\to
2\tau$ \\
\hline
$3.8\sigma$ & $2\sigma$ & $1.4\sigma$ & $1.1\sigma$ & $14\sigma$ \\
\hline
\end{tabular}
\end{center}
\vspace*{-.2in}
\end{table}
}

To further illustrate the nature of the Higgs sector for a $\mueff>0$
scenario, I give details of a sample $\tanb=40$ inverted-ideal Higgs
point in Table~\ref{pt10p}.  Let me make a few remarks.  For this
somewhat unusual point, the $\hi$ is fairly singlet, the $\hii$ and
$\aii$ are largely $H_d$ and it is the $\hiii$ that is most
SM-like. Nonetheless, the effective precision electroweak mass
(defined by $\ln \meff\equiv \sum_{i=1,2,3} [g_{ZZh_i}^2/g_{ZZ
  \hsm}^2]\ln m_{h_i}$) receives substantial contributions from the
low mass Higgses, $\hi$ and $\hii$, and lies below $105\gev$ and is
thus in the range that is ideal for precision electroweak data.  Next,
note that Higgs decays to $\cnone\cnone$ are unimportant, but that
Higgs to Higgs pair decays are often significant.  Prospects for LHC
detection of some of the Higgs are very good.  In particular, $b\anti
b\hii+b\anti b\aii$ with $\aii,\hii\to \tauptaum$ should be readily
observable.  Even $\ai$ discovery via $gg\to \ai \to \mupmum$ looks
promising~\cite{Dermisek:2009fd} because $\caibb\sim 6$ and $\mai$ is
not directly under the $\Upsilon_{3S}$ peak. The SM-like $\hiii$ is
possibly the most difficult to detect because: a) its $\gam\gam$ decay
mode is suppressed by $C_V(1)<1$ and mass $\mhiii\sim 126\gev$ that is
above the mass where the branching ratio to $\gam\gam$ is maximal;
and, b) its decays to Higgs pairs will reduce all standard detection
modes.

Finally, let me note that in the NMSSM study of \cite{Das:2010ww}
cross sections as large as those found here are not achieved; rather,
they find $\sig^{Max}_{SI}\sim (1-1.5)\times 10^{-6}\pb$ (without
enhancing the $s$-quark content of the nucleon).  The smaller
$\sig_{SI}$ is largely because they did not seek scenarios with
$\hi\sim H_d$.  Ref.~\cite{Das:2010ww} also considers the possibility
of enhancing $\sigsi$ by a factor of $\sim 3$ by enhancing the
$s$-quark content of the nucleon. If we adopted this same enhancement,
our $\sigsi$ values would approach the lowered \cogent/DAMA $\sigsi$ range
applicable if we also employed the higher local density of $\rho\sim
0.485\gev/{\rm cm}^2$ mentioned earlier.  However, current analyses do
not appear to allow for such a large $s$-quark enhancement~\cite{leutwyler}.

\section{The Singlino-Singlet Scenarios}

In these SS scenarios, the LSP is primarily singlino and the Higgs
responsible for large $\sigsi$ is mainly singlet.
In~\cite{Belikov:2010yi}, we pursued the extended NMSSM as defined
earlier and looked for scenarios of the SS type. What we found was a
kind of see-saw balance between $\Omega_{\cnone} h^2$ and $\sigsi$ such that
when $\Omega_{\cnone} h^2\sim 0.1$ then $\sigsi$ is very naturally in the
\cogent/DAMA preferred zone.  Below, I provide a few details.

The singlino coupling to down-type quarks is
given by:
\begin{equation}
\frac{a_d}{m_d} =  \frac{g_2 \kappa N^2_{15} \tan\beta F_s(\hi) F_d(\hi)}{8 m_W m^2_{h_1}}
\end{equation}
where $\hi = F_d(\hi) H^0_d + F_u(\hi) H_u^0 + F_s(\hi) H_S^0$ and the
crucial trilinear coupling that couples a singlino pair to the singlet
Higgs $H_S^0$ is proportional to $\kap$. This leads to
\begin{eqnarray}\label{eq:sigma}
\sigsi &\approx& 2.2 \times 10^{-4}\pb   \bigg(\frac{\kappa}{0.6}\bigg)^2 \, \bigg(\frac{\tan \beta}{50}\bigg)^2 \bigg(\frac{45\, {\rm GeV}}{m_{h_1}}\bigg)^4  \bigg(\frac{F^2_s(\hi)}{0.85}\bigg) \bigg(\frac{F^2_d(\hi)}{0.15}\bigg), \nonumber
\end{eqnarray}
which is consistent with the value required by \cogent\ and DAMA/LIBRA
for the indicated $\kappa$, $\mhi$ and $\hi$ component values.  (Of
course, one really sums coherently over all the CP-even Higgs bosons.)
Furthermore, the large singlet fraction $F^2_s(\hi) \sim 0.85$ of the
$h_1$ will allow it evade the constraints from LEP II and
the Tevatron.

Meanwhile, the thermal relic density of neutralinos is 
determined by the annihilation cross section and the $\cnone$ mass. In
the mass range we are considering here, the dominant annihilation
channel is to $b\bar{b}$ (or, to a lesser extent, to $\tau^+ \tau^-$)
through the $s$-channel exchange of 
the {\em same} scalar Higgs, $h_1$, as employed for elastic
scattering, yielding:
\begin{eqnarray}
\sigma_{\chi_1^0\chi_1^0} v &=& 
\frac{N_c g^2_2 \kappa^2 m^2_b F^2_s(\hi) F^2_d(\hi)}{64 \pi m^2_W  \cos^2\beta}
\frac{m^2_{\chi_1^0} (1-m^2_b/m^2_{\chi_1^0})^{3/2} \, v^2}
{(4 m^2_{\chi_1^0}-m^2_{h_1})^2 + m^2_{h_1} \Gamma^2_{h_1}},~~~
\label{eq:omega}
\end{eqnarray}
where $v$ is relative velocity between the annihilating neutralinos,
$N_c = 3$ is a color factor and $\Gamma_{h_1}$ is the width 
of the exchanged Higgs. The annihilation cross section into $\tau^+ \tau^-$ is obtained 
by replacing $m_b \rightarrow m_{\tau}$ and $N_c \rightarrow 1$. 
This yields the thermal relic abundance of neutralinos:
%
%\begin{equation}
$
\Omega_{\chi_1^0} h^2 \approx \frac{10^9}{M_{\rm Pl}}\frac{m_{\chi_1^0}}{T_{\rm FO} \sqrt{g_{\star}}}
\frac{1}{\langle \sigma_{\chi_1^0 \chi_1^0} v \rangle},
$
%\end{equation}
%
where $g_{\star}$ is the number of relativistic degrees of freedom at
freeze-out, $\langle \sigma_{\chi_1^0 \chi_1^0} v \rangle$ is the
thermally averaged annihilation cross section at freeze-out, and
$T_{\rm FO}$ is the temperature at which freeze-out occurs.

For the range of masses and cross sections considered here, we find 
$m_{\chi_1^0}/T_{\rm FO}\approx 20$, yielding a thermal relic abundance of
\begin{eqnarray}
\Omega_{\chi_1^0} h^2 
\approx 0.11 \,  \bigg(\frac{0.6}{\kappa}\bigg)^2 \bigg(\frac{50}{\tan \beta}\bigg)^2 \bigg(\frac{m_{h_1}}{45 \, {\rm GeV}}\bigg)^4 
 \bigg(\frac{7 \, {\rm GeV}}{m_{\chi_1^0}}\bigg)^2
\bigg(\frac{0.85}{F^2_s(\hi)}\bigg) \bigg(\frac{0.15}{F^2_d(\hi)}\bigg),
\label{eq:abundance}
\end{eqnarray}
\ie\ naturally close to the measured dark matter density, $\Omega_{\rm
  CDM} h^2 = 0.1131 \pm 0.0042$ {\em for the same choices for
  $\kappa$, $\mhi$ and composition fractions as give \cogent/DAMA-like
  $\sigsi$}.  The only question is can we achieve the above
situation without violating LEP and other constraints. Basically, one
wants a certain level of decoupling between the singlet sectors and
the MSSM sectors, but not too much.  To find out, we performed
parameter scans with an extended version of NMHDECAY and micrOMEGAs
that includes both the non-NMSSM parameters of Eqs. (\ref{eq:W}) and
(\ref{eq:Lsoft}) as well as the latest $B$-physics and Tevatron
constraints.  We find points for $15<\tan\beta<45$ that are consistent
(within the usual $\pm 2\sigma$ combined theory plus experimental
windows -- excursions in $b\to s\gamma$ and $b\bar b
h,h\to\tau^+\tau^-$ that fall slightly outside this window
 are present at high $\tanb$)
with all collider and $B$-physics constraints having the appropriate
thermal relic density and $\sigsi$ as large as $few
\times 10^{-4}\pb$.

 The complete framework has contributions to
$\sigsi$ and $\Omega_{\chi_1^0}h^2$ beyond
Eqs. (\ref{eq:sigma}) and (\ref{eq:abundance}) and
high-$\sigsi$ points typically have large contributions
from the non-singlet Higgses.  I confine myself to discussing one
'typical' point that does the job. Its properties are tabulated 
in Table~\ref{pt18348}.

{\tiny 
\begin{table}[t!]
\vspace*{-.3in}
  \caption{\small Properties of a typical ENMSSM point with
    $\tanb=45$ and $\msusy=1000\gev$.\label{pt18348}}
\vspace*{-.1in}
\begin{center}
\begin{tabular}{|c|c|c|c|c|c|c|c|c|}
\hline
$\lam$ & $\kap$ & $\lam s$ & $\alam$ & $\akap$ & $M_1$ & $M_2$ & $M_3$ & $\asoft$
\cr
\hline
$0.011$ &   $0.596$ & $ -0.026\gev$ & $3943\gev $ & $17.3\gev$ & $150\gev$ &
$300\gev$ & $900\gev$ & $679\gev$ \cr
\hline
\end{tabular}

\begin{tabular}{|c|c|c|c|c|c|c|}
\hline
$B_S$ & $\mu_S$ &  $v_S^3$ & $\mu$ & $B_\mu$ & $\mueff$ & $B_\mu^{eff}$ 
\cr
\hline
$0$ &   $7.8\gev$ & $ 4.7 \gev$ & $164\gev $ & $658\gev$ &
$164\gev$ & $556\gev$  \cr
\hline
\end{tabular}

\begin{tabular}{|c|c|c|c|c|c|}
\hline
$\mhi$ & $\mhii$ & $\mhiii$ & $\mai$ & $\maii$ & $\mhp$ \cr
\hline
$82\gev$ & $118\gev$ & $164\gev$ & $82\gev$ & $164\gev$ &
$178\gev$  \cr 
\hline
\end{tabular}

\begin{tabular}{|c|c|c|c|c|c|c|c|}
\hline
 $F_S^2(\hi)$& $F_d^2(\hi)$ & $F_S^2(\hii)$ & $F_u^2(\hii)$ &
 $F_S^2(\hiii) $ & $ F_d^2(\hiii)$ &  $F_S^2(\ai)$ & $F_S^2(\aii)$ \cr
\hline
$0.86$ & $0.14$ & $0.0$ & $0.996$ & $0.14$ & $0.86$ & $0.86$ & $0.14$ \cr
\hline
\end{tabular}

\begin{tabular}{|c|c|c|c|c|c|c|c|}
\hline
$C_V(\hi)$ & $C_V(\hii)$ & $C_V(\hiii)$ & $\chibb$ & $\chiibb$ & $\chiiibb$ & $\caibb$ & $\caiibb$ \cr
\hline
 $-0.0096$ &  $0.999$ & $ -0.041$ & $16.8$ & $2.9$ & $41.7$ & $-16.9$ & $41.7$ \cr
\hline
\end{tabular}

\begin{tabular}{|c|c|c|c|c|c|c|c|}
\hline
$\mcnone$ & $N_{11}^2$ & $N_{13}^2+M_{14}^2$ & $N_{15}^2$  & $\sigsi$  & $\Omega_{\cnone} h^2$  \cr 
\hline
$4.9\gev$ & $0.0$ & $0.0$ & $1.0$  & $2.0 \times 10^{-4} \pb$ &  $0.105$ \cr
\end{tabular}

\begin{tabular}{|c|c|c|c|c|}
\hline  
$\br(\hi\to\cnone\cnone)$ & $\br(\hi\to 2b,2\tau)$ &
$\br(\hii\to\cnone\cnone$ & $\br(\hii\to
2b,2\tau)$ & $\br(\hp\to \tau^+\nu) $
 \cr 
\hline
   $0.64$ & $0.33,0.03$ & $0.003$ & $0.88,0.092$ & $0.97$ \cr 
  \hline
\end{tabular}

\begin{tabular}{|c|c|c|c|}
\hline
$\br(\ai\to\cnone\cnone)$ & $\br(\ai\to 2b,2\tau)$ &
$\br(\aii,\hiii\to\cnone\cnone)$ & $\br(\aii,\hiii\to 2b,2\tau)$ \cr
\hline
 $0.64$ & $0.33,0.03$ &   $0.05$ & $0.85,0.095$ \cr
\hline
\end{tabular}
 \end{center}
\vspace*{-.3in}
\end{table}
}
Let us note the following regarding this particular point.
\ben
\item What you see is that the $\hi,\ai$ have separated off from
  something that is close to an MSSM-like Higgs sector with $\hii\sim\hl$
  being SM-like and $\hiii\sim \hh$, $\aii \sim \ha$ and $\hp\sim
  H^+$.

\item Detection of the $\hii$ would be possible via the usual SM-like
  detection modes planned for the MSSM $\hl$.
  
\item There are some $\hii,\aii\to \cnone\cnone$ decays, but at such a
  low branching ratio level that detection of these invisible decay
  modes would be unlikely, even if very interesting.

\item Decays to pairs of Higgs of any of the heavier Higgs bosons are
  not of importance. Of course, by choosing $\msusy=1000\gev$ so that
  $\mhii>114\gev$ (beyond the LEP limits), we have not forced the
  issue.  It will be interesting to look for SS scenarios that are
  ideal-Higgs-like with $\mhii<110\gev$. 

\item One sees that $\hi$ and $\ai$ decay primarily to $\cnone\cnone$
  but that there also decays to $b\anti b$ and $\tauptaum$ with
  reduced branching ratios of $0.33$ and $0.03$ compared to the normal
  $\br(b\anti b)\sim 0.85$ and $\br(\tauptaum)\sim 0.12$.

\item $\hi$ and $\ai$ do have somewhat enhanced couplings to $b\anti
  b$ (in this example $\chibb,\caibb\sim \sqrt{F_d^2(\hi,\ai)}\tanb\sim 17$)
  and so the rates for $gg\to b\anti b \hi+gg\to b\anti b \ai$ will be
  quite substantial. However, the reduced $\br(\hi,\ai\to
  \tauptaum)\sim 0.03$ implies that detection of such production in
  the $b\anti b+\tauptaum$ final state might prove challenging,
  probably requiring very high $L$ at the LHC.

\item Further work is needed to quantify discovery prospects in the
  $gg \to b\anti b +(\hi,\ai)\to b\anti b +\etmiss$ channel.

\item At this large $\tanb$, detection of the $\hiii$ and $\aii$ would
  certainly be possible in $gg\to b\anti b \hiii+b\anti b\aii$ in the
  $\hiii,\aii\to\tauptaum$ decay channel.

\item For this sample case, the charged Higgs is {\it just} too heavy to allow
  $t\to \hp b$ decays and so one would have to turn to $gg\to \anti t
  b \hp+t \anti b \hm$ with detection of the charged Higgs in the
  $\tau\nu_\tau$ final state.  Further investigation is needed to
  assess the feasibility of such detection, but at least the cross
  section is very enhanced by virtue of the large $\tanb$ value.

  \een A few final notes regarding this scenario.  First, it is the
  very large value of $\alam$ and the very small $\lam$ that keep
  the singlet and MSSM Higgs sectors fairly separate.  Second, 
  the new parameters of the ENMSSM, $\mu$
  and $B_\mu$ must be substantial.  This is generally the case if you
  desire an SS scenario with  $\mhi>few\gev$.
  
{\tiny 
\begin{table}[h!]
\vspace*{-.3in}
  \caption{\small Properties of the SS DLH NMSSM point
    with $\tanb=13.77$, $m_{\wtil q}=1000\gev$ and $m_{\wtil \ell}=200\gev$.\label{dlh}}
\vspace*{-.1in}
\begin{center}
\begin{tabular}{|c|c|c|c|c|c|c|c|c|}
\hline
$\lam$ & $\kap$ & $\lam s$ & $\alam$ & $\akap$ & $M_1$ & $M_2$ & $M_3$ & $\asoft$
\cr
\hline
$0.1205$ &   $0.00272$ & $ 168\gev$ & $2661\gev $ & $-24.03\gev$ & $100\gev$ &
$200\gev$ & $660\gev$ & $750\gev$ \cr
\hline
\end{tabular}

\begin{tabular}{|c|c|c|c|c|c|}
\hline
$\mhi$ & $\mhii$ & $\mhiii$ & $\mai$ & $\maii$ & $\mhp$ \cr
\hline
$0.811\gev$ & $116\gev$ & $244\gev$ & $16.7\gev$ & $244\gev$ &
$244\gev$  \cr 
\hline
\end{tabular}

\begin{tabular}{|c|c|c|c|c|c|c|c|}
\hline
 $F_S^2(\hi)$& $F_d^2(\hi)$ & $F_S^2(\hii)$ & $F_u^2(\hii)$ &
 $F_S^2(\hiii) $ & $ F_d^2(\hiii)$ &  $F_S^2(\ai)$ & $F_S^2(\aii)$ \cr
\hline
$0.997$ & $0.00017$ & $0.0036$ & $0.99$ & $0.0$ & $0.994$ & $1.00$ & $0.00$ \cr
\hline
\end{tabular}

\begin{tabular}{|c|c|c|c|c|c|c|c|}
\hline
$C_V(\hi)$ & $C_V(\hii)$ & $C_V(\hiii)$ & $\chibb$ & $\chiibb$ & $\chiiibb$ & $\caibb$ & $\caiibb$ \cr
\hline
 $0.06$ &  $0.998$ & $ 0.0$ & $0.183$ & $0.994$ & $13.77$ & $-0.12$ & $13.77$ \cr
\hline
\end{tabular}

\begin{tabular}{|c|c|c|c|c|c|c|c|}
\hline
$\mcnone$ & $N_{11}^2$ & $N_{13}^2+N_{14}^2$ & $N_{15}^2$  & $\sigsi$  & $\Omega_{\cnone} h^2$  \cr 
\hline
$7.2\gev$ & $0.0036$ & $0.017$ & $0.98$  & $2.34 \times 10^{-4} \pb$ &
$0.112$ \cr
\hline
\end{tabular}

\begin{tabular}{|c|c|c|c|c|}
\hline  
$\br(\hi\to\cnone\cnone)$ & $\br(\hi\to 2s,2g,2\mu)$ &
$\br(\hii\to\cnone\cnone)$ & $\br(\hii\to \cnone\cntwo)$ & $\br(\hii\to
2b,2\tau)$ 
 \cr 
\hline
   $0.027$ & $0.833,0.14,0.027$ & $0.05$ & $0.45$ & $0.37,0.038$  \cr 
  \hline
\end{tabular}

\begin{tabular}{|c|c|}
\hline
$\br(\hp\to t \bar b) $ & $\br(\hp\to \wtil \chi^+_{1,2}\wtil\chi^0_{1,2,3,4,5})$ \cr
\hline
$0.138$ & $0.80$ \cr
\hline
\end{tabular}

\begin{tabular}{|c|c|}
\hline
$\br(\ai\to\cnone\cnone)$ & $\br(\ai\to 2b,2\tau,2\mu)$ 
\cr
\hline
 $0.25$ & $0.70,0.042,0.00015$ \cr
\hline
\end{tabular}
\begin{tabular}{|c|c|c|c|}
\hline
$\br(\aii,\hiii\to\cnone\cnone)$ & $\br(\aii,\hiii\to 2t,2b,2\tau)$ &
$\br(\aii,\hiii\to \wtil \chi^0_{1,2,3,4,5}\wtil \chi^0_{1,2,3,4,5}) $ 
& $\br(\aii,\hiii\to \wtil \chi^+_{1,2}\wtil \chi^-_{1,2}) $ \cr
\hline
$0.00$ & $0.013,0.126,0.023$ & $0.32$ & $0.48$ \cr
\hline
\end{tabular}
 \end{center}
\vspace*{-.2in}
\end{table}
}

  As noted earlier, in~\cite{Draper:2010ew} an alternative SS scenario
  can be realized in the strict NMSSM, but only if $\mhi\lsim 1\gev$.
  The properties of their representative point are tabulated in
  Table~\ref{dlh}. Some observations regarding this scenario are the
  following.
\ben
\item The $\hi$ is very light and very singlet. It is so weakly
  coupled to the down and up quarks that it can probably only be
  detected directly via $\Upsilon_{3S}\to \gam \hi$ with $\hi\to
  \mupmum$. For current data from BaBar and using
  $\br(\hi\to\mupmum)\sim 0.027$ (see the Table), the limit from
  $\Upsilon_{3S}\to\gam\hi\to \gam \mupmum$ is $\chibb\sim 0.2-0.3$
  for $\mhi\sim 1\gev$ (the limit fluctuates very rapidly). For this
  scenario the value of $\chibb=0.183$ (see the Table) is thus just
  below the BaBar limit.  This indicates that increased statistics
  could very well reveal the light $\hi$ since $\chibb$ cannot be much
  below this value and still provide a large enough $\sigsi$ to
  explain the \cogent/DAMA events.
\item
  Meanwhile, the $\hii$ is completely SM-like and its
  discovery at the LHC or Tevatron would be possible in the usual
  channels for a SM Higgs of the same mass.  

\item 

  The $\ai$ has a very small branching ratio to $\mupmum$ (since
  $\mai>2m_B$) and would have to be searched for in the $b\anti b$ or
  $\tauptaum$ decay mode.  Since the $\ai$ is very singlet its
  production cross sections would be so small that this would likely be
  an impossible task.

\item The $\hiii,\aii,\hp$ form a decoupled degenerate doublet with
  common mass of around $244\gev$. The $b\anti b \hiii$ and $b\anti b
  \aii$ couplings are both enhanced by a factor of $\tanb=13.77$.  The
  most promising LHC signal would be $gg\to b\anti b \hiii+ b\anti b
  \aii$ with decay $\hiii,\aii\to \tauptaum$.  Of course,
  $\br(\hiii,\aii\to \tauptaum)\sim 0.023$ is uncomfortably small and
  even this signal would be quite weak (and does not emerge in the
  NMHDECAY LHC estimates as viable).

\een

\section{Conclusions}

The \cogent/DAMA data suggests a large spin-independent cross section
for dark matter scattering on nucleons, $\sigsi\sim (1-3)\times
10^{-4}\pb$, at low dark matter mass, $m_{DM}\sim 4-9\gev$.  This
cannot be achieved for the lightest neutralino in the minimal
supersymmetric model (MSSM) after imposing all constraints, including,
in particular, $\Omega_{\cnone}h^2 \sim 0.11$ and the Tevatron limit
of $\br(B_s\to \mupmum)<5.8\times 10^{-8}$. It is then natural to ask
if the simplest extension of the MSSM obtained by adding a singlet
superfield to the MSSM (the NMSSM and ENMSSM models) can allow
simultaneous compatibility between \cogent/DAMA events and all other
constraints, or must one turn to more exotic supersymmetric or other
models.  As reviewed here, the NMSSM and ENMSSM {\em can} achieve
large $\sigsi$ at low $\mcnone$ while satisfying all constraints, but
only if the Higgs sector has the appropriate structure and properties.
Indeed, given the LEP, Tevatron, BaBar and other constraints, only a
limited number of possibilities within the NMSSM and ENMSSM have been
delimited to date.  These include:

\ben
\item  The ``inverted-Higgs'' (IH)
scenarios where the lightest CP-even Higgs, $\hi$, is $H_d$-like while the
$\hii$ has SM-like couplings to $WW,ZZ$, but might decay via
$\hii\to\ai\ai$. 

After imposing $\Omega_{\cnone}h^2\sim 0.11$, and all other
constraints, the value of $\sigsi$ that can be achieved falls short by
something like a factor of $5-10$ (assuming a reasonable $(g-2)_\mu$ is
required) in comparison to the value of $\sigsi\sim (1-3)\times
10^{-4}\pb$ that is apparently required by \cogent/DAMA. However, this
kind of scenario could become interesting if: a) the relevant $\sigsi$
values at small $\mcnone$ turn out to be somewhat smaller (\eg\ new
data or increased local density $\rho$); b) the $(g-2)_\mu$
restrictions employed turn out to be incorrect; and/or c) the
$s$-quark content of the nucleons has been underestimated.

All Higgs bosons in the IH scenarios are quite light and discovery
prospects are good. The $\cnone$ is primarily bino. One can even have
an ``inverted-ideal-Higgs'' (IIH) scenario in which the $ZZ$ coupling
squared weighted Higgs mass, $\meff$, is below $105\gev$ and is thus
in the ideal range for precision electroweak data, finetuning and
electroweak baryogenesis. We focused on describing
such an IIH scenario in our discussion.

\item The singlet-singlino (SS) scenarios in which the $\hi$ and
  $\cnone$ are primarily singlet and singlino, respectively, with
  $\mhi$ fairly small ($\mhi\sim 40-70\gev$) in the ENMSSM case and
  very small ($\mhi\lsim 1\gev$) in the NMSSM
  DLH case.   

In both these cases, one can achieve the required
  $\sigsi\sim 2\times 10^{-4}\pb$ while maintaining
  $\Omega_{\cnone}h^2\sim 0.11$ and obeying all constraints.  

In the ENMSSM case,  detecting the $\hi$ directly at the
  LHC in $gg\to b\anti b\hi$ with $\hi\to\tauptaum$ might prove possible.

  In the DLH case, detection of the $\hi$ would require a significant
  (but not enormous) increase in statistics relative to current BaBar
  data for $\Upsilon_{3S}\to\gam\mupmum$.

  In both cases, most of the other Higgs bosons would be readily
  detectable.  

\een 

In general, there is an intimate connection between achieving large
$\sigsi$ at small $\mcnone$ and a relatively unusual Higgs sector
structure. In the SS cases, detection of the singlet $\hi$ will be
highly non-trivial, but absolutely necessary if we are to understand
the source of the large $\sigsi$ and achieve a quantitative
understanding of $\Omega_{\cnone} h^2$. If at least some Higgs bosons
are discovered and their properties measured and if the \cogent/DAMA
events are confirmed as dark matter detection, then it is also
possible that the NMSSM and ENMSSM supersymmetric models will be ruled
out, requiring that more exotic possibilities be considered.

\section{Acknowledgments}

This work was supported by US DOE grant DE-FG03-91ER40674. I wish to
thank my collaborators for their contributions to our joint work. I
also wish to thank the organizers of PASCOS 2010 for their generous
support and the Aspen Center for Physics for support during the course
of this project. I also wish to thank C. Wagner for several useful
communications.

\end{document}